\documentclass[preprint,aps,showpacs]{revtex4}

\usepackage{graphicx}
\usepackage{bm}       
\usepackage{mathptmx} 

\begin{document}

\title{Ambiguities in the up quark mass}
\author{Michael Creutz}
\affiliation{
Physics Department, Brookhaven National Laboratory\\
Upton, NY 11973, USA
}

\begin{abstract}
{ It has long been known that no physical singularity is encountered
as up quark mass is adjusted from small positive to negative values as
long as all other quarks remain massive.  This is tied to an additive
ambiguity in the definition of the quark mass.  This calls into
question the acceptability of attempts to solve the strong CP problem
via a vanishing mass for the lightest quark.}
\end{abstract}

\maketitle

The standard SU(3) non-Abelian gauge theory of the strong interactions
is quite remarkable in that, once an arbitrary overall scale is fixed,
the only parameters are the quark masses.  Using a few pseudo-scalar
meson masses to fix these parameters, the non-Abelian gauge theory
describing quark confining dynamics is unique.

When multiple quark masses vanish, the theory acquires exact chiral
symmetries.  These manifest themselves through Goldstone bosons
arising from spontaneous symmetry breaking in the vacuum.  As the
physical light quarks do have masses, these symmetries are only
approximate, but the pions are believed to be the remnants of this
Goldstone boson structure.

The case where only one of the quarks is massless is particularly
interesting in that anomalies break all chiral symmetries.  A mass gap
is dynamically generated, no Goldstone bosons are expected, and no
singularities occur as the mass passes through zero.  Because of the
anomalies, negative and positive quark masses are not equivalent.
Indeed, when a single quark mass is made sufficiently negative a
spontaneous breakdown of parity will appear \cite{dashen, mymasspaper,
paper1}.

The smooth behavior around small quark masses raises the question of
how to define the quark mass.  Because of confinement, it is not
obvious how experiments involving only physical particles can
determine whether a non-degenerate quark mass vanishes.  If this
cannot be done, then the concept of a massless quark is unphysical.

Whether the lightest quark is massless might be regarded as an
academic point.  Phenomenologically, despite ambiguities in the chiral
Lagrangian \cite{Kaplan:1986ru}, it appears that a massless up quark
seems untenable \cite{Gerard:1989mr,Leutwyler:1996qg}.  Nevertheless,
even in principle the mass is just some parameter adjusted to give the
correct long distance physics.  From this point of view a vanishing
value is just a point in parameter space, as good as any other.  But
the issue gains significance when a special value of the parameter is
used to solve a more complex problem.  This is the case with the up
quark mass, which continues to be proposed as a possible solution to
what is known as the strong CP problem \cite{pecceiquinn, banks}.  

The purpose of this paper is to emphasize how confinement and chiral
anomalies make it unclear whether the concept of a vanishing $m_u$ is
well posed.  Ref. \cite{banks} raises some of these issues, pursuing
$m_u=0$ anyway as an accidental symmetry.  A preliminary unpublished
version of these arguments is contained in Ref.~\cite{Creutz:2003cj}.

Because renormalization is required, the concept of an ``underlying
basic Lagrangian'' does not exist.  The continuum theory is specified
in terms of basic symmetries and a few renormalized parameters.  In
practice, the definition of a field theory relies on a limiting
process from a cutoff version.  As the lattice is the best understood
non-perturbative cutoff, it provides the most natural framework for
such a definition.  Thus I denote my cutoff parameter as $a$,
representing a lattice spacing or minimum length.

This is only a notational issue.  Any regulator must accommodate the
known chiral anomalies, and thus chiral symmetry breaking terms of
some form must appear in the cutoff theory.  These effects can come in
many guises.  With a Pauli-Villars scheme, there is a heavy regulator
field.  With dimensional regularization the anomaly is hidden in the
fermionic measure.  For Wilson lattice gauge theory there is the
famous Wilson term.  With domain wall fermions there is a residual
mass from a finite fifth dimension.  With overlap fermions things are
hidden in a combination of the measure and a certain non-uniqueness of
the operator.  A lattice regulator also introduces a dimensionful
parameter, the lattice spacing $a$.  This feature also is not special
to the lattice.  The scale anomaly, through the phenomenon known as
``dimensional transmutation'' \cite{cw}, is responsible for masses of
hadrons such as the proton and glueballs, even in the massless quark
limit.  For such physics, any complete regulator must introduce a
scale.

For the issue being raised here, the heavier quarks play no crucial
role.  Thus I imagine them to be ``integrated out'' and consider the
theory reduced to a single flavor.  This allows me to consider only
two bare parameters, the coupling and quark mass.  Including the other
quarks is straightforward, but unnecessarily complicates the
equations.

The renormalization process tunes all relevant bare parameters as a
function of the cutoff while fixing a set of renormalized quantities.
As I need to renormalize both the bare coupling and quark mass, I need
to fix two physical observables.  For this purpose I choose the
lightest boson and the lightest baryon masses.  As both are expected
to be stable, this precludes any ambiguity from particle widths.  From
its roots in the multi-flavor theory, I denote the lightest boson the
$\pi$, and the lightest baryon as $p$.  Because of confinement, the
values of their masses are inherently non-perturbative quantities.

With the cutoff in place, the physical masses are functions of
$(g,m,a)$, the bare charge, the bare coupling, and the cutoff.
Holding the masses constant, the renormalization process determines
how $g$ and $m$ flow as the cutoff is removed.  Because of asymptotic
freedom, this flow eventually enters the perturbative regime and we
have the famous renormalization group equations \cite{Gell-Mann:fq}
\begin{eqnarray}
a{dg\over da}&\equiv&\beta(g)=\beta_0 g^3+\beta_1 g^5 +\ldots \\
a{dm\over da}&\equiv&m\gamma(g)=m(\gamma_0 g^2+\gamma_1 g^4 +\ldots)
+{\rm non\hbox{-}perturbative}.
\end{eqnarray}
The ``non-perturbative'' term should vanish faster than any power of
the coupling.  I include it explicitly in the mass flow because it
will play a crucial role in the latter discussion.  The values for the
first few coefficients $\beta_0$, $\beta_1$, and $\gamma_0$ are known
\cite{coef} and independent of regularization scheme.

The solution to these equations shows how the bare coupling and bare
mass are driven to zero as the cutoff is removed
\begin{eqnarray}
a&=&{1\over \Lambda} e^{-1/2\beta_0 g^2} g^{-\beta_1/\beta_0^2}
(1+O(g^2))\label{flow1}\\
m&=&Mg^{\gamma_0/\beta_0}
(1+O(g^2)).
\label{flow2}
\end{eqnarray}
The quantities $\Lambda$ and $M$ are integration constants for the
renormalization group equations.  I refer to $\Lambda$ as the overall
strong interaction scale and $M$ as the renormalized quark mass.
Their values depend on the explicit renormalization scheme as well as
the physical values being held fixed in the renormalization process,
i.e. the proton and pion masses.  This connection is highly
non-perturbative.  Indeed particle masses are long distance
properties, and thus require following the renormalization group flow
far out of the perturbative regime.

Turning things around, we can consider the physical particle masses to
be functions of these integration constants.  Simple dimensional
analysis tells us that the dependence of physical masses must take the
form
\begin{eqnarray}
m_p&=&\Lambda f_p(M/\Lambda)\\
m_\pi&=&\Lambda f_\pi(M/\Lambda)
\end{eqnarray}
where the $f_i(x)$ are dimensionless functions whose detailed form is
highly non-perturbative.

For the case of degenerate quarks we expect the square of the pion
mass to vanish linearly as the renormalized quark mass goes to zero.
This means we anticipate a square root singularity in $f_\pi(x)$ at
$x=0$.  Indeed, requiring the singularity to occur at the origin
removes any additive non-perturbative ambiguity in defining the
renormalized mass.  

With a non-degenerate quark, things are more subtle.  As discussed
above, we expect physics to behave smoothly as the quark mass passes
through zero.  That is, we do not expect $f_i(x)$ to display any
singularity at $x=0$.  Non-perturbative dynamics generates an
additional contribution to the mass of the pseudo-scalar meson; thus,
the $M=0$ flow generically corresponds to a positive value of $m_\pi$.
While a $m_\pi=0$ flow line can exist, it represents the boundary of a
CP violating phase and has little to do with massless quarks.

Now I come to the question of scheme dependence.  Given some different
renormalization prescription, i.e. a modified lattice action, the
precise flows will change.  Although the behavior dictated in
Eqs.~(\ref{flow1},\ref{flow2}) will be preserved, the integration
constants $(\Lambda,M)$ and the function $f(x)$ will in general be
modified.  Marking the new quantities with tilde's, matching the
schemes to give the same physics requires
\begin{equation}
m_i=\Lambda f_i(M/\Lambda)=\tilde \Lambda 
\tilde f_i(\tilde M/\tilde\Lambda).
\end{equation}
Upon the removal of the cutoff, two different valid cutoff schemes
should give the same result for the physical masses.  If the concept
of a massless quark has physical meaning, this means that $M=0$ should
match with $\tilde M=0$.  Otherwise the continuum limit in one scheme
for the massless quark theory would correspond to the continuum limit
taken in another scheme where the quark mass is not zero.  The issue
raised in this paper is the absence of any known reason for the
vanishing of $M$ to require the vanishing of $\tilde M$.

\begin{figure*}
\centering
\includegraphics[width=2.5in]{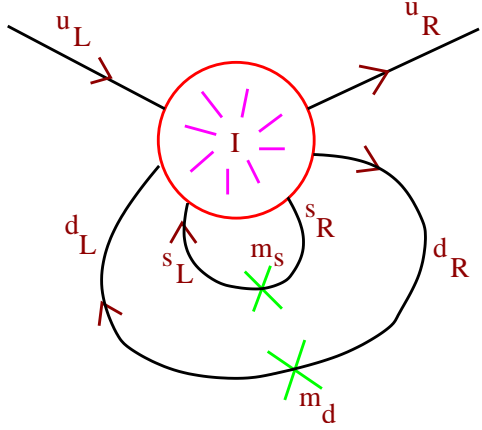}
\caption{
\label{instanton}
Non-perturbative classical gauge configurations will generate an
effective mass term for the up quark.  The magnitude of this mass is
proportional to the product of the heavier quark masses.  The precise
value, however, is scheme and scale dependent.}
\end{figure*} 

On changing schemes, we introduce new definitions for the coupling and
mass.  To match onto the perturbative limit, it is reasonable to
restrict these definitions to agree at leading order.  Thus I require
\begin{eqnarray}
\tilde g&=& g+O(g^3)\label{pertmatch1}\\
\tilde m&=& m(1+O(g^2))+{\rm non\hbox{-}perturbative.}
\label{pertmatch2}
\end{eqnarray}
Here the ``non-perturbative'' terms should vanish faster than any
power of the coupling, but are not in general proportional to $m$.  In
particular, a non-perturbative additive shift in the up-quark mass
follows qualitatively from the analysis of classical gauge
configurations, i.e. ``pseudo-particles'' or ``instantons''
\cite{mcarthurgeorgi}.  As shown some time ago by 't Hooft
\cite{thooft}, these configurations generate an effective
multi-fermion vertex where all flavors of quark flip their spin.  If
we take this vertex and tie together the massive quark lines with mass
terms, then the resulting process generates an effective mass term for
the light quark.  This process is illustrated in
Fig.~(\ref{instanton}).  The strength of this term is proportional to
the product of the masses of the more massive quarks.  If there are no
additional massive quarks, the scale is set by $\Lambda$, the strong
interaction scale.

The requirements for the perturbative limit apply at fixed cutoff.
Indeed, the interplay of the $a\rightarrow 0$ and the $g\rightarrow 0$
limits is rather intricate.  As $g\rightarrow 0$ at fixed $a$ the
quarks decouple and we have a theory of free quarks and gluons.  The
limit $a\rightarrow 0$ at fixed $g$ brings on the standard divergences
of relativistic field theory.  The proper continuum limit follows the
renormalization group trajectory with both $a$ and $g$ going together
in the appropriate way and gives a theory with important
non-perturbative effects such as confinement.

Assuming only the matching conditions in
Eq.~(\ref{pertmatch1},\ref{pertmatch2}) leaves the freedom to do some
amusing things.  As a particularly contrived example, consider
\begin{eqnarray}
\tilde g &=& g\\
\tilde m&=&m-M g^{\gamma_0/\beta_0}\times
{ e^{-1/2\beta_0 g^2} g^{-\beta_1/\beta_0^2}\over \Lambda a}.
\end{eqnarray}
The last factor vanishes than any power of $g$, but is crafted to go
to unity along the renormalization group trajectory.  Note that a
power of the scale factor is necessary for non-perturbative phenomena
to be relevant to the continuum limit \cite{thooft}.  With this form,
one can immediately relate the old and new renormalized masses
\begin{equation}
\tilde M\equiv
\lim_{a\rightarrow 0}\ \tilde m \tilde g^{-\gamma_0/\beta_0} = 
M-M=0.
\end{equation}
Thus for any $M$, another scheme always exists where the renormalized
quark mass vanishes.  The possibility of such a transformation is the
root of the claim that masslessness is not a physical concept for a
non-degenerate quark.

Now I turn to some observations on the relevance of this conclusion to
lattice gauge theory.  Recently there has been considerable progress
with lattice fermion formulations that preserve a remnant of exact
chiral symmetry \cite{myreview}.  With such, the multiple flavor
theory will preserve the $m_\pi=0$ contour as the $m=0$ axis.  The
important point is that this is not true for the one flavor theory,
where the $m_\pi=0$ flow delves into the negative mass regime.  The
motivation for extending these chiral fermion actions to the one
flavor case seems extremely perverse; indeed, in this situation we do
not expect any exact chiral symmetry to survive.  But if we are to
give a massless quark any scheme independent meaning, this may be the
only route.  Nevertheless, even with these actions, there is still no
reason to expect $M=0$ to give scheme independent physical masses.

To begin with, the chiral fermion actions are not themselves unique.
For example, the overlap operator \cite{overlap} is constructed by a
projection process from the conventional Wilson lattice operator.  The
latter has a mass parameter which is to be chosen in a particular
domain.  On changing this parameter, the massless Dirac operator still
satisfies the Ginsparg-Wilson relation \cite{gw}, but this condition
does not guarantee that physical particle masses shift in a way that
preserves their ratio.

As another way to see that this non-universality might be expected,
consider that the dynamics of the one flavor case generates a mass gap
in the $\pi$ channel.  This means that the eigenvalues of the Dirac
operator important to low energy physics are are not near the origin,
but dynamically driven a finite distance away \cite{Banks:1979yr}.
Indeed, the absence of massless particles in the regulated theory
requires the density of these eigenvalues to vanish at the origin.
Changing the projection procedure generating the overlap operator will
modify the size of this gap, changing the $\pi$ mass.  If the baryon
mass is not equally affected by exactly the same factor, this
modification cannot be absorbed in an overall scale factor.  The
effects on the baryon mass, however, are expected to be less dependent
on chiral issues since the baryon remains massive even when several
quarks are massless.

Finally, the action used for the pure gauge field is not unique.  It
has been demonstrated that different gauge actions can strongly modify
the topological structures at finite lattice spacing
\cite{Aoki:2002vt}.  The effect of such structures is expected to
dominate the boson mass generation but play a lesser role for the
baryon.

This non-perturbative ambiguity in the quark mass carries over to the
explicitly CP violating case where the mass is complex with phase
$\theta$.  The above discussion shows that even the sign of the up
quark mass can be ambiguous.  Thus different schemes can have an
ambiguity of $\theta$ between 0 and $\pi$, a particularly severe
example.  For other values of $\theta$, to fix the continuum theory
uniquely we need to introduce another renormalized quantity.  For
example, this could be a three meson coupling or the electric dipole
moment of a baryon.  Beyond the lowest order in the chiral expansion,
the precise dependence of the renormalized parameter on $\theta$ is
scheme-dependent.

Although I have phrased the discussion in terms of a mass parameter,
the conclusions are unchanged if the mass is generated via a Higgs
mechanism involving unifying fields.  The Yukawa coupling of the Higgs
field to the up quark receives an additive shift induced by the
heavier quarks interacting with non-perturbative gauge fields.  This
shift gives a vanishing Yukawa coupling a significance similar to a
vanishing fundamental quark mass.

In summary, I have argued that the concept of a single massless quark
is mathematically ill posed.  Admittedly this is not a rigorous proof.
But any conclusion depending fundamentally on the concept of a
massless up quark, such as the preservation of CP symmetry in unified
theories, should address how the miracle of a vanishing $M$ gives
scheme independent physical particle masses.

\section*{Acknowledgements}
I have benefited from lively discussions with many colleagues, in
particular F. Berruto, T. Blum, K. Petrov, H. Neuberger,
G. Senjanovic, and Y. Shamir.  This manuscript has been authored under
contract number DE-AC02-98CH10886 with the U.S.~Department of Energy.
Accordingly, the U.S. Government retains a non-exclusive, royalty-free
license to publish or reproduce the published form of this
contribution, or allow others to do so, for U.S.~Government purposes.


\begin{thebibliography}{99}
\bibitem{dashen}
R.~F.~Dashen,
Phys.\ Rev.\ D {\bf 3}, 1879 (1971).

\bibitem{mymasspaper}
M.~Creutz,
Phys.\ Rev.\ D {\bf 52}, 2951 (1995)
[arXiv:hep-th/9505112].

\bibitem{paper1}
M.~Creutz,
arXiv:hep-lat/0312018 (2003).

\bibitem{cw}
S.~R.~Coleman and E.~Weinberg,
Phys.\ Rev.\ D {\bf 7} (1973) 1888.

\bibitem{Kaplan:1986ru}
D.~B.~Kaplan and A.~V.~Manohar,
Phys.\ Rev.\ Lett.\  {\bf 56}, 2004 (1986).

\cite{Gerard:1989mr}
\bibitem{Gerard:1989mr}
J.~M.~Gerard,
Mod.\ Phys.\ Lett.\ A {\bf 5}, 391 (1990).

\bibitem{Leutwyler:1996qg}
H.~Leutwyler,
Phys.\ Lett.\ B {\bf 378}, 313 (1996)
[arXiv:hep-ph/9602366].

\bibitem{pecceiquinn}
R.~D.~Peccei and H.~R.~Quinn,
Phys.\ Rev.\ Lett.\  {\bf 38}, 1440 (1977);
H.~Leutwyler,
Nucl.\ Phys.\ B {\bf 337}, 108 (1990);
A.~G.~Cohen, D.~B.~Kaplan and A.~E.~Nelson,
JHEP {\bf 9911}, 027 (1999)
[arXiv:hep-lat/9909091];
D.~R.~Nelson, G.~T.~Fleming and G.~W.~Kilcup,
Phys.\ Rev.\ Lett.\  {\bf 90}, 021601 (2003);
R.~J.~Crewther, P.~Di Vecchia, G.~Veneziano and E.~Witten,
Phys.\ Lett.\ B {\bf 88}, 123 (1979)
[Erratum-ibid.\ B {\bf 91}, 487 (1980)].

\bibitem{banks}
T.~Banks, Y.~Nir and N.~Seiberg,
arXiv:hep-ph/9403203.

\bibitem{Creutz:2003cj}
M.~Creutz,
arXiv:hep-th/0303254 (2003, unpublished).

\bibitem{Gell-Mann:fq}
M.~Gell-Mann and F.~E.~Low,
Phys.\ Rev.\  {\bf 95}, 1300 (1954);
K.~Symanzik,
Commun.\ Math.\ Phys.\  {\bf 18}, 227 (1970);
C.~G.~.~Callan,
Phys.\ Rev.\ D {\bf 2}, 1541 (1970).

\bibitem{coef}
H.~D.~Politzer,
Phys.\ Rev.\ Lett.\  {\bf 30}, 1346 (1973);
D.~J.~Gross and F.~Wilczek,
Phys.\ Rev.\ Lett.\  {\bf 30}, 1343 (1973);
Phys.\ Rev.\ D {\bf 8}, 3633 (1973);
W.~E.~Caswell,
Phys.\ Rev.\ Lett.\  {\bf 33}, 244 (1974);
D.~R.~T.~Jones,
Nucl.\ Phys.\ B {\bf 75}, 531 (1974);
J.~A.~M.~Vermaseren, S.~A.~Larin and T.~van Ritbergen,
Phys.\ Lett.\ B {\bf 405}, 327 (1997)
[arXiv:hep-ph/9703284];
K.~G.~Chetyrkin,
Phys.\ Lett.\ B {\bf 404}, 161 (1997)
[arXiv:hep-ph/9703278].

\bibitem{mcarthurgeorgi}
H.~Georgi and I.~
N.~McArthur,
Harvard University preprint HUTP-81/A011 (1981, unpublished).

\bibitem{thooft}
G.~'t Hooft,
Phys.\ Rev.\ D {\bf 14}, 3432 (1976)
[Erratum-ibid.\ D {\bf 18}, 2199 (1978)].

\bibitem{myreview}
M.~Creutz,
Rev.\ Mod.\ Phys.\  {\bf 73}, 119 (2001)
[arXiv:hep-lat/0007032].

\bibitem{overlap}
H.~Neuberger,
Phys.\ Lett.\ B {\bf 417}, 141 (1998)
[arXiv:hep-lat/9707022].

\bibitem{gw}
P.~H.~Ginsparg and K.~G.~Wilson,
Phys.\ Rev.\ D {\bf 25}, 2649 (1982).

\bibitem{Banks:1979yr}
T.~Banks and A.~Casher,
Nucl.\ Phys.\ B {\bf 169}, 103 (1980).

\bibitem{Aoki:2002vt}
Y.~Aoki {\it et al.},
arXiv:hep-lat/0211023.

\end{thebibliography}
\end{document}